\documentstyle[preprint,aps]{revtex}
\begin{document}
\draft
\title{Exact Results for the Adsorption of a Flexible
Self-Avoiding Polymer Chain in Two Dimensions}
\author{M. T. Batchelor and C. M. Yung}
\address{Department of Mathematics, School of Mathematical Sciences,
Australian National University, Canberra ACT 0200, Australia}
\date{\today}
\maketitle
\begin{abstract}
We derive the exact critical couplings ($x^*, y_{\rm a}^*$), where
$y_{\rm a}^*/x^* = \sqrt{1+\sqrt2} = 1.533\ldots\,$, for the polymer
adsorption transition on the honeycomb lattice, along with the
universal critical exponents, from the Bethe Ansatz solution of
the O($n$) loop model at the special transition.
Our result for the
thermal scaling dimension, and thus the crossover exponent
$\phi=\frac{1}{2}$, is in agreement with an earlier result based
on conformal invariance arguments. Our result for the geometric
scaling dimensions confirms recent conjectures that they are given
by $h_{\ell+1,3}$ in the Kac formula.
\end{abstract}

\narrowtext

A long flexible polymer in a good solvent with an attractive
short-range force between the polymer and the container wall is known
to undergo an adsorption transition\cite{deg,ekb,htw,dl}.
A standard model for this phenomenon is a
self-avoiding walk (SAW) on a $d$-dimensional lattice interacting with
a ($d-1$)-dimensional substrate.
In the lattice model, the SAW has a
Boltzmann weight $x$ per monomer (in the bulk), with weight
$y$ per adsorbed monomer (on the substrate).
At the adsorption transition, $y_{\rm a}^*$, the number of adsorbed monomers
scales with the total length $L$ as $L_{\rm a} \sim L^\phi$, where $\phi$
is a crossover exponent.
The polymer is in the adsorbed phase for $y > y_{\rm a}^*$ and the
desorbed phase for $y < y_{\rm a}^*$, where the surface attractions are
not effective.
In the language of surface critical phenomena, the adsorption
transition is a {\it special} transition\cite{bin}.

Two-dimensional polymers are not without experimental interest\cite{exp}
and the above model has been widely studied via a number of
techniques (see, e.g., \cite{dl} and references therein).
These include transfer-matrix calculations\cite{gb},
series expansions\cite{zld} and a scanning Monte Carlo method\cite{mc}.
In two dimensions there is a wealth of exact results for the
{\it ordinary} surface transition ($y < y_{\rm a}^*$) from conformal
invariance arguments\cite{c,bc,ds,dl} and more recently from
exact Bethe Ansatz calculations\cite{bs}.
More generally, these results have been obtained for the
O($n$) model, from which the configurational properties of
SAWs follow in the $n \rightarrow 0$ limit\cite{deg}.

The situation is not so clear for the special transition.
There is a conformal invariance result for the thermal
scaling dimension $X_\epsilon$, which leads to the crossover exponent
$\phi = \frac{1}{2}$\cite{beg}.
However, the two-variable nature of the special
transition poses problems for numerical studies, as
errors in the estimates of the critical exponents are compounded
by errors in the location of the critical point ($x^*, y_{\rm a}^*$).
Recent simulations on the square lattice have indicated a result
significantly larger than the conjectured $\phi$ value\cite{mc}.

Here we derive the exact critical couplings
($x^*, y_{\rm a}^*$) for the polymer adsorption transition on
the honeycomb lattice, along with the universal critical exponents,
from the Bethe Ansatz solution of the O($n$) loop model at the
special transition.

Our starting point is the partition function of an O($n$) loop model
\cite{dmns} defined on the honeycomb lattice depicted in Fig. 1,
\begin{equation}
{\cal Z}_{{\rm loop}} = \sum x^{L} y^{L_s} n^P,
\end{equation}
where the sum is over all configurations of closed and nonintersecting
loops. Here $P$ is the total number of closed loops of fugacity $n$
in a given configuration. In the limit $n\rightarrow0$
this reduces to the required SAW generating function,
with $x$ the fugacity of a step in the bulk and $y$ the fugacity
of a step along the surface. Here
$L$ is the length of a walk in the bulk and $L_s$ is the length
of a walk along the surface of the strip.

The partition function can be conveniently rewritten in terms of the
Boltzmann weights of the empty vertices. To do this we need to distinguish
between three classes of vertices: $\rangle\!-$ and
$-\!\langle$ which appear (i) in the bulk and (ii)
on the surface, and (iii) $\rangle$ and $\langle$ on the surface.
For each class we define the weights
$t_b$, $t_{\bar b}$ and $t_s$, respectively. We then consider
\begin{equation}
{\cal Z}_{{\rm loop}} = \sum t_b^{{\cal N}_b-L_b}
           t_{\bar b}^{{\cal N}_{\bar b}-L_{\bar b}-{1\over2}L_s}
                             t_s^{{\cal N}_s-{1\over2}L_s} n^P,
\end{equation}
where ${\cal N}_b$, ${\cal N}_{\bar b}$ and ${\cal N}_s$ are the total
numbers of vertices (either full or empty) of class (i), (ii) and (iii).
Apart from harmless normalisation factors, the two partition functions
are equivalent if $t_b = t_{\bar b}$, along with the identification
\begin{equation}
x = 1/t_b, \quad y = 1/\sqrt{t_{\bar b} t_s}
\label{couplings}
\end{equation}
where $L = L_b + L_{\bar b}$.

The configurations of the loop model can be mapped to those of a 3-state
vertex model in the standard way\cite{n1,b1,n3}. The allowed arrow
configurations
and their corresponding Boltzmann weights are shown in Fig. 2. Here the
phase factors are such that $n = s + s^{-1} = -2 \cos 4\lambda$. The
integrable bulk weights of this honeycomb lattice model are known to
follow in a particular limit of the Izergin-Korepin model -- a more
general 3-state model defined on the square lattice\cite{ik,n3}.

For the open boundary conditions of interest here, the
integrability of the vertex model can be examined in a systematic
way by making use of reflection or $K$-matrices which satisfy the boundary
version of the Yang-Baxter equation. In order to do this, we adapted the
Sklyanin construction of commuting transfer matrices\cite{skl} to
the present geometry\cite{dd,yb,foot1}.
In particular, we found that the known diagonal reflection matrices for the
Izergin-Korepin model\cite{mn} lead to two integrable
sets of boundary weights which preserve the O($n$) symmetry\cite{yb,foot2}.
For each case $t_b=t_{\bar b}=2 \cos \lambda$.
Thus from (\ref{couplings}) the critical bulk fugacity is
\begin{equation}
1/x^*=\sqrt{2 \pm \sqrt{2-n}},
\end{equation}
which is the well known bulk critical value\cite{n1}.
The SAW point occurs at $\lambda = \pi/8$, where
$x^*=1/\sqrt{2 + \sqrt{2}}=0.541~196\ldots$.

The two integrable sets of boundary weights are\cite{yb}
\begin{equation}
\mbox{(A)}\quad t_s = \frac{\sin 2 \lambda}{\sin \lambda}\quad\mbox{and}
\quad \mbox{(B)}\quad t_s = \frac{\cos 2 \lambda}{\cos \lambda}.
\end{equation}
Thus from (\ref{couplings}) case (A) gives the critical surface fugacity
$y_{\rm o}^* = x^*$ and corresponds to an integrable point on the
ordinary transition line.
However case (B) is new, and corresponds to the special transition,
with
\begin{equation}
y_{\rm a}^*=\left(2-n\right)^{-1/4}, \label{cc}
\end{equation}
in the so-called dilute phase $0 \le \lambda \le \pi/4$ ($-2 \le n \le 2$).
In contrast the surface coupling is complex-valued in the
dense phase ($\pi/4 < \lambda \le \pi/2$). We thus confine our attention
here to the dilute region applicable to the adsorption transition.
At $n=0$ we have
$y_{\rm a}^* = 2^{-1/4}=0.840~896\ldots$ This exact result
should prove to be
a valuable benchmark for future numerical studies of the adsorption
transition. For $n=1$ the special transition is located at
$y_{\rm a}^* = 1$ which corresponds to infinitely strong surface
couplings, as expected. Our result is also consistent with
a recent argument that although there is no special transition for
the Ising model, there {\it is} a special transition in the
geometrical O($n$) model for $n \ge 1$\cite{fs}.
We see from (\ref{cc}) that the critical coupling diverges at $n=2$,
and becomes complex for $n > 2$.

The central charge $c$ and scaling dimensions $X_i$ defining the critical
behaviour of the model follow from the dominant finite-size corrections
to the transfer matrix eigenvalues\cite{c1,c2,cx}.
The central charge follows from the
free energy per site, $f_N = N^{-1} \log \Lambda_0$, via
\begin{equation}
f_N \simeq f_{\infty} + \frac{f_s}{N} + \frac{\pi \zeta c}{6 N^2}.
\label{cf}
\end{equation}
Here $f_s$ is the surface free energy and
$\zeta=\sqrt 3 /2$ is a lattice-dependent scale factor.
The scaling dimensions are related to the inverse
correlation lengths via
\begin{equation}
\xi_i^{-1} = \log ( \Lambda_0/\Lambda_i) \simeq 2 \pi \zeta X_i/N.
\label{Xf}
\end{equation}

A set of scaling dimensions of interest appear in the so-called
watermelon correlator, which measures the geometric correlation
between $\ell$ nonintersecting SAWs tied together at their
extremities $\mbox{\boldmath $x$}$ and $\mbox{\boldmath $y$}$, which for
surface critical phenomena are near the boundary
of the half-plane\cite{ds}.  It has a critical algebraic decay,
\begin{equation}
G_\ell(\mbox{\boldmath $x$}-\mbox{\boldmath $y$}) \sim
|\mbox{\boldmath $x$}-\mbox{\boldmath $y$}|^{-2 X_\ell}.
\end{equation}
These scaling dimensions are associated with the largest eigenvalue
in each sector of the transfer matrix. In particular, the
spin-spin correlation is related to $X_1$.
On the other hand, the thermal scaling dimension $X_\epsilon$,
corresponding to the energy-energy correlation, is
related to an excitation in the largest sector of the transfer matrix.
Given $X_1$ and $X_\epsilon$, the
surface critical exponents follow from\cite{bin,dl}
\begin{eqnarray}
X_1 &=& \eta_{\parallel}/2 = \beta_1/\nu = [2 + (\gamma-2\gamma_1)/\nu]/2,
\label{f1}\\
X_\epsilon &=& 1 - \phi/\nu. \label{f2}
\end{eqnarray}
In terms of the more standard parameter $g$, where $4\lambda + \pi g = 2\pi$,
the bulk exponents $\gamma$ and $\nu$ are given by\cite{n1}
\begin{equation}
\nu = \frac{g}{4(g-1)}, \qquad \gamma = \left(
\mbox{\small $\frac{3}{4}$}\, g + 1/g \right) \nu . \label{f3}
\end{equation}

For case (B), at the special transition $(x^*, y_{\rm a}^*)$, we have obtained
the Bethe Ansatz solution for the eigenvalues of the transfer matrix\cite{yb},
\begin{equation}
\Lambda = \prod_{j=1}^{m}
{    {\sinh(u_j+ {\rm i}\, 3\lambda/2) \sinh(u_j-{\rm i}\, 3\lambda/2)}
                     \over
     {\sinh(u_j+{\rm i}\, \lambda/2) \sinh(u_j-{\rm i}\,\lambda/2)} },
\end{equation}
where the $u_j$ follow as roots of the Bethe Ansatz equations
\widetext
\begin{eqnarray}
\lefteqn{\Bigl[
  {{\cosh(u_j-{\rm i}\,\lambda/2)}
  \over{\cosh(u_j+{\rm i}\,\lambda/2)}}\Bigr]^2
 \Bigl[
  {{\sinh(u_j-{\rm i}\,\lambda/2) \sinh(u_j-{\rm i}\,3\lambda/2)}
\over{\sinh(u_j+{\rm i}\,\lambda/2)} \sinh(u_j+{\rm i}\,3\lambda/2)}
\Bigr]^N =} \nonumber\\
& & \quad \prod_{\stackrel{k=1}{\ne j}}^{m}
{ {\sinh(u_j-u_k+{\rm i}\,\lambda) \sinh(u_j+u_k+{\rm i}\,\lambda)
   \sinh(u_j-u_k-{\rm i}\,2\lambda) \sinh(u_j+u_k-{\rm i}\,2\lambda) }
         \over
  {\sinh(u_j-u_k-{\rm i}\,\lambda) \sinh(u_j+u_k-{\rm i}\,\lambda)
   \sinh(u_j-u_k+{\rm i}\,2\lambda) \sinh(u_j+u_k+{\rm i}\,2\lambda)  }  }.
\end{eqnarray}
\narrowtext\noindent
Here $N$ is the width of the strip (e.g., $N=8$ in Fig. 1) and
$m$ labels the sectors of the transfer matrix, with
$m=N$ for the largest eigenvalue $\Lambda_0$. A more convenient
sector label is $\ell = N - m$.

The Bethe Ansatz equations differ from those obtained for case (A) in the
squared prefactor on the left hand side, which have the $\cosh$ functions
replaced by $\sinh$. This change is sufficient to alter the finite-size
corrections to the eigenvalues and thus the operator content.
For case (A), at the ordinary transition, the
Bethe Ansatz roots for the largest eigenvalue are uniformly distributed
along the real (positive) axis\cite{bs}.
In contrast, at the special transition we find that the root distribution
includes an elementary 1-string excitation, located at
$u_1 \sim {\rm i} \left( \frac{\pi}{4} + \frac{\lambda}{2} \right)$.
To derive the central charge via (\ref{cf}) we thus adopt the analytic
method \cite{wbn}, which avoids the explicit manipulation of root densities.
Details of the calculations will be presented elsewhere.
The bulk free energy $f_\infty$ is as derived
previously\cite{b1,bs}, while the surface free energy $f_s$ differs
from that at the ordinary transition\cite{bs}.
The result is cumbersome and rather unilluminating in the present
context. However, at $n=0$, it reduces to
$f_s = - 2 \log (1+\sqrt 2)$. In fact at this point we observe that
$\Lambda_0 = (2+\sqrt 2 )^N/(1+\sqrt 2)^2$ exactly.

For the central charge, we derive the same result,
\begin{equation}
c = 1 - 6 (g-1)^2/g,
\end{equation}
as for the ordinary transition\cite{c,ds,bs}.
We find that the thermal scaling dimension is associated
with an elementary 2-string excitation, with
\begin{equation}
X_\epsilon = \frac{2}{g} -1, \label{Xt}
\end{equation}
in agreement with the conformal invariance result\cite{beg}. This result has
more recently been obtained from a thermodynamic Bethe Ansatz calculation
from a conjectured boundary $S$-matrix \cite{fs}.
{}From (\ref{f2}) and (\ref{f3}) the result (\ref{Xt}) leads to the
crossover exponent $\phi = \mbox{\small $\frac{1}{2}$}$.

The root distributions for the eigenvalues defining the
the geometric scaling dimensions are again real, with
\begin{equation}
X_\ell = {\mbox{\small $\frac{1}{4}$}}g (\ell+1)^2 -
{\mbox{\small $\frac{3}{2}$}}(\ell+1) +
\frac{9-(g-1)^2}{4 g}, \label{Xg}
\end{equation}
where $\ell = 1,2,\ldots$ These dimensions are to be compared with
those at the ordinary transition, where the result\cite{ds,bs}
\begin{equation}
X_\ell = {\mbox{\small $\frac{1}{4}$}}g \ell^2 +
{\mbox{\small $\frac{1}{2}$}}(g-1) \ell,
\end{equation}
follows from $h_{\ell+1,1}$ in the Kac formula\cite{ds},
\begin{equation}
h_{p,q} = {\mbox{\small $\frac{1}{4}$}} g p^2 -
{\mbox{\small $\frac{1}{2}$}} p q +
\frac{q^2-(g-1)^2}{4 g}.
\end{equation}

For the adsorption transition, (\ref{Xg}) gives
\begin{equation}
X_\ell = {\mbox{\small $\frac{3}{8}$}}\ell(\ell-2) +
{\mbox{\small $\frac{1}{3}$}}
\end{equation}
at $n=0$ ($g=\mbox{\small $\frac{3}{2}$})$.
The first two values are $X_1 = - \mbox{\small $\frac{1}{24}$}$ and
$X_2 = X_\epsilon = \mbox{\small $\frac{1}{3}$}$, with the
exact exponents for the two-dimensional polymer adsorption transition
following from (\ref{f1}) and (\ref{f2}).
In particular, $\eta_{\parallel}= - \mbox{\small $\frac{1}{12}$}$ and
the susceptibility exponent $\gamma_1 = \mbox{\small $\frac{93}{64}$}$.
Guim and Burkhardt\cite{gb} originally noted
that these were possibly the exact values, as
their finite-size scaling estimates for $X_1$ and $X_2$ were
compatible with $X_\ell = h_{\ell+1,3}$ in the Kac formula.
Indeed we confirm that the more general O($n$) result (\ref{Xg})
agrees with $h_{\ell+1,3}$.
Note that the thermal dimension (\ref{Xt}) also belongs to the same family
of scaling dimensions, since $X_\epsilon = h_{1,3}$.
We believe that our results exhaust the complete operator content of the
O($n$) model at the special transition.

More recently the $h_{\ell+1,3}$ result has been conjectured to be
correct by Fendley and Saleur\cite{fs} who
argued that the boundary operator $\Phi_{\ell+1,3}$ propagates
down the strip at the special point.
In general, our
exact results lend further weight to their claim that the spin degrees of
freedom of the Kondo problem can be considered as the $n=2$ limit of
the special transition of the O($n$) model\cite{fs}.

So far we have only considered the O($n$) model with boundary
conditions that are symmetric with the left and right boundaries
of the strip.
More generally it is possible to obtain the Bethe ansatz solutions
of the O($n$) model with non-symmetric boundary conditions by using
suitable choices of the reflection matrices. In this
way we expect to test recent conformal invariance results for the
O($n$) model with mixed boundary conditions\cite{be}.

\acknowledgments

It is a pleasure to thank A.~L. Owczarek and J. Suzuki for helpful comments.
This work has been supported by the Australian Research Council.

\newpage

\begin{figure}
\vskip 2cm
\setlength{\unitlength}{0.0030000in}%
\begingroup\makeatletter
\def\x#1#2#3#4#5#6#7\relax{\def\x{#1#2#3#4#5#6}}%
\expandafter\x\fmtname xxxxxx\relax \def\y{splain}%
\ifx\x\y   
\gdef\SetFigFont#1#2#3{%
  \ifnum #1<17\tiny\else \ifnum #1<20\small\else
  \ifnum #1<24\normalsize\else \ifnum #1<29\large\else
  \ifnum #1<34\Large\else \ifnum #1<41\LARGE\else
     \huge\fi\fi\fi\fi\fi\fi
  \csname #3\endcsname}%
\else
\gdef\SetFigFont#1#2#3{\begingroup
  \count@#1\relax \ifnum 25<\count@\count@25\fi
  \def\x{\endgroup\@setsize\SetFigFont{#2pt}}%
  \expandafter\x
    \csname \romannumeral\the\count@ pt\expandafter\endcsname
    \csname @\romannumeral\the\count@ pt\endcsname
  \csname #3\endcsname}%
\fi
\endgroup
\begin{center}
\begin{picture}(669,735)(161,79)
\thicklines
\put(708,294){\line( 1, 2){ 27}}
\put(737,347){\line(-3, 5){ 30.441}}
\put(707,398){\line(-1, 0){ 60}}
\put(647,398){\line(-3,-5){ 31.323}}
\put(617,345){\line( 3,-5){ 30.441}}
\put(647,294){\line( 1, 0){ 61}}
\put(800,245){\line( 3, 5){ 30.882}}
\put(830,297){\line(-3, 5){ 31.147}}
\put(799,349){\line(-1, 0){ 60}}
\put(739,348){\line(-3,-5){ 30.882}}
\put(709,296){\line( 3,-5){ 31.147}}
\put(740,244){\line( 1, 0){ 60}}
\put(800,140){\line( 3, 5){ 31.323}}
\put(830,193){\line(-3, 5){ 31.147}}
\put(799,245){\line(-1, 0){ 60}}
\put(739,244){\line(-3,-5){ 30.882}}
\put(709,192){\line( 3,-5){ 31.147}}
\put(740,140){\line( 1, 0){ 60}}
\put(619,139){\line( 3, 5){ 31.147}}
\put(650,191){\line(-3, 5){ 30.706}}
\put(619,242){\line(-1, 0){ 60}}
\put(559,242){\line(-3,-5){ 31.588}}
\put(528,189){\line( 3,-5){ 30.971}}
\put(560,138){\line( 1, 0){ 59}}
\put(525,293){\line( 3, 5){ 30.882}}
\put(555,345){\line(-3, 5){ 31.147}}
\put(524,397){\line(-1, 0){ 60}}
\put(464,396){\line(-3,-5){ 30.882}}
\put(434,344){\line( 3,-5){ 30.441}}
\put(464,293){\line( 1, 0){ 61}}
\put(796,449){\line( 3, 5){ 31.323}}
\put(826,502){\line(-3, 5){ 30.441}}
\put(796,553){\line(-1, 0){ 61}}
\put(735,552){\line(-3,-5){ 30.882}}
\put(705,500){\line( 3,-5){ 30.706}}
\put(736,449){\line( 1, 0){ 60}}
\put(614,449){\line( 3, 5){ 31.588}}
\put(645,502){\line(-3, 5){ 31.147}}
\put(614,554){\line(-1, 0){ 60}}
\put(554,553){\line(-3,-5){ 30.882}}
\put(524,501){\line( 3,-5){ 31.147}}
\put(555,449){\line( 1, 0){ 59}}
\put(796,656){\line( 3, 5){ 30.618}}
\put(825,708){\line(-3, 5){ 30.882}}
\put(795,760){\line(-1, 0){ 60}}
\put(735,759){\line(-3,-5){ 30.882}}
\put(705,707){\line( 3,-5){ 30.441}}
\put(735,656){\line( 1, 0){ 61}}
\put(705,604){\line( 3, 5){ 31.323}}
\put(735,657){\line(-3, 5){ 30.441}}
\put(705,708){\line(-1, 0){ 61}}
\put(644,708){\line(-3,-5){ 30.882}}
\put(614,656){\line( 3,-5){ 31.147}}
\put(645,604){\line( 1, 0){ 60}}
\put(524,604){\line( 3, 5){ 30.882}}
\put(554,656){\line(-3, 5){ 31.147}}
\put(523,708){\line(-1, 0){ 60}}
\put(463,707){\line(-3,-5){ 30.882}}
\put(433,655){\line( 3,-5){ 30.882}}
\put(463,603){\line( 1, 0){ 61}}
\put(431,448){\line( 3, 5){ 31.147}}
\put(462,500){\line(-3, 5){ 31.147}}
\put(431,552){\line(-1, 0){ 60}}
\put(371,551){\line(-3,-5){ 30.882}}
\put(341,499){\line( 3,-5){ 31.147}}
\put(372,447){\line( 1, 0){ 59}}
\put(344,290){\line( 3, 5){ 30.618}}
\put(373,342){\line(-3, 5){ 30.882}}
\put(343,394){\line(-1, 0){ 61}}
\put(282,394){\line(-1,-2){ 27}}
\put(253,341){\line( 3,-5){ 30.882}}
\put(283,289){\line( 1, 0){ 61}}
\put(436,136){\line( 3, 5){ 30.618}}
\put(465,188){\line(-3, 5){ 30.882}}
\put(435,240){\line(-1, 0){ 61}}
\put(374,240){\line(-1,-2){ 27}}
\put(345,187){\line( 3,-5){ 30.882}}
\put(375,135){\line( 1, 0){ 61}}
\put(254,133){\line( 3, 5){ 30.882}}
\put(284,185){\line(-3, 5){ 30.882}}
\put(254,237){\line(-1, 0){ 61}}
\put(193,236){\line(-3,-5){ 30.882}}
\put(163,184){\line( 3,-5){ 31.147}}
\put(194,132){\line( 1, 0){ 60}}
\put(254,342){\line( 3, 5){ 30.618}}
\put(283,394){\line(-3, 5){ 30.882}}
\put(253,446){\line(-1, 0){ 61}}
\put(192,445){\line(-3,-5){ 30.618}}
\put(163,393){\line( 3,-5){ 30.882}}
\put(193,341){\line( 1, 0){ 61}}
\put(252,446){\line( 3, 5){ 31.323}}
\put(282,499){\line(-3, 5){ 31.147}}
\put(251,551){\line(-1, 0){ 60}}
\put(191,550){\line(-3,-5){ 31.323}}
\put(161,497){\line( 3,-5){ 30.706}}
\put(192,446){\line( 1, 0){ 60}}
\put(341,603){\line( 3, 5){ 31.147}}
\put(372,655){\line(-3, 5){ 30.706}}
\put(341,706){\line(-1, 0){ 60}}
\put(281,706){\line(-3,-5){ 31.323}}
\put(251,653){\line( 3,-5){ 30.441}}
\put(281,602){\line( 1, 0){ 60}}
\put(252,656){\line( 3, 5){ 30.882}}
\put(282,708){\line(-3, 5){ 30.706}}
\put(251,759){\line(-1, 0){ 60}}
\put(191,759){\line(-3,-5){ 31.323}}
\put(161,706){\line( 3,-5){ 30.706}}
\put(192,655){\line( 1, 0){ 60}}
\put(345,288){\line( 3,-5){ 28.412}}
\put(466,290){\line(-3,-5){ 30}}
\put(527,292){\line( 2,-3){ 33.231}}
\put(556,344){\line( 1, 0){ 63}}
\put(647,292){\line(-1,-2){ 24.800}}
\put(434,447){\line( 3,-5){ 28.412}}
\put(342,600){\line( 2,-3){ 30.923}}
\put(524,604){\line( 3,-5){ 28.853}}
\put(706,603){\line( 3,-5){ 28.853}}
\put(616,447){\line( 3,-5){ 28.412}}
\put(797,446){\line( 3,-5){ 28.412}}
\put(797,653){\line( 3,-5){ 28.412}}
\put(163,284){\line( 3,-5){ 28.853}}
\put(163,599){\line( 3,-5){ 28.853}}
\put(432,757){\line( 2,-3){ 30.923}}
\put(613,756){\line( 3,-5){ 28.853}}
\put(286,185){\line( 1, 0){ 62}}
\put(469,187){\line( 1, 0){ 62}}
\put(649,190){\line( 1, 0){ 62}}
\put(374,342){\line( 1, 0){ 62}}
\put(463,499){\line( 1, 0){ 63}}
\put(282,498){\line( 1, 0){ 62}}
\put(370,654){\line( 1, 0){ 61}}
\put(554,656){\line( 1, 0){ 62}}
\put(647,501){\line( 1, 0){ 63}}
\put(282,286){\line(-3,-5){ 29.471}}
\put(374,448){\line(-3,-5){ 30.618}}
\put(556,450){\line(-3,-5){ 30.177}}
\put(737,448){\line(-3,-5){ 30.882}}
\put(829,401){\line(-3,-5){ 30.177}}
\put(828,607){\line(-3,-5){ 30.177}}
\put(645,603){\line(-3,-5){ 30.618}}
\put(463,600){\line(-3,-5){ 30.441}}
\put(283,604){\line(-3,-5){ 30.177}}
\put(191,653){\line(-3,-5){ 30.177}}
\put(193,338){\line(-3,-5){ 30.441}}
\put(372,760){\line(-3,-5){ 30.177}}
\put(554,759){\line(-3,-5){ 30.441}}
\put(194,130){\line(-3,-5){ 30.177}}
\put(375,132){\line(-3,-5){ 30.177}}
\put(559,134){\line(-3,-5){ 30.882}}
\put(742,139){\line(-3,-5){ 30.441}}
\put(282,811){\line(-3,-5){ 30.618}}
\put(463,811){\line(-3,-5){ 30.618}}
\put(643,807){\line(-3,-5){ 30.177}}
\put(255,129){\line( 3,-5){ 28.853}}
\put(436,134){\line( 3,-5){ 29.118}}
\put(621,136){\line( 3,-5){ 28.853}}
\put(343,809){\line( 3,-5){ 28.412}}
\put(524,809){\line( 2,-3){ 30.923}}
\put(706,807){\line( 3,-5){ 28.853}}
\put(373,759){\line( 1, 0){ 63}}
\put(554,758){\line( 1, 0){ 62}}
\put(163,807){\line( 3,-5){ 28.853}}
\put(800,139){\line( 3,-5){ 28.853}}
\put(826,814){\line(-3,-5){ 30.177}}
\end{picture}
\end{center}

\vskip 2cm
\caption{The open honeycomb lattice.}
\end{figure}
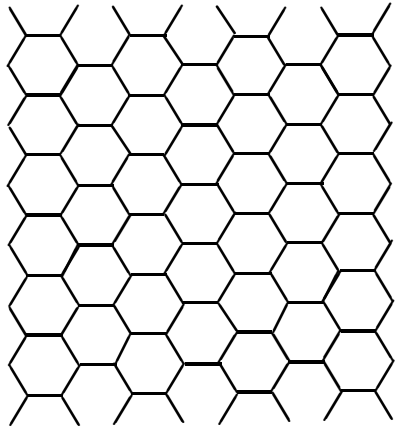

\newpage

\begin{figure}
\setlength{\unitlength}{0.00400in}%
\begingroup\makeatletter
\def\x#1#2#3#4#5#6#7\relax{\def\x{#1#2#3#4#5#6}}%
\expandafter\x\fmtname xxxxxx\relax \def\y{splain}%
\ifx\x\y   
\gdef\SetFigFont#1#2#3{%
  \ifnum #1<17\tiny\else \ifnum #1<20\small\else
  \ifnum #1<24\normalsize\else \ifnum #1<29\large\else
  \ifnum #1<34\Large\else \ifnum #1<41\LARGE\else
     \huge\fi\fi\fi\fi\fi\fi
  \csname #3\endcsname}%
\else
\gdef\SetFigFont#1#2#3{\begingroup
  \count@#1\relax \ifnum 25<\count@\count@25\fi
  \def\x{\endgroup\@setsize\SetFigFont{#2pt}}%
  \expandafter\x
    \csname \romannumeral\the\count@ pt\expandafter\endcsname
    \csname @\romannumeral\the\count@ pt\endcsname
  \csname #3\endcsname}%
\fi
\endgroup
\begin{center}
\begin{picture}(908,577)(37,149)
\thicklines
\put(129,725){\line( 3,-5){ 32.735}}
\put(161,670){\line(-3,-5){ 32.471}}
\put(164,670){\line( 1, 0){ 56}}
\put(336,324){\line(-3,-5){ 32.735}}
\put(304,269){\line( 3,-5){ 32.471}}
\put(713,216){\line( 3, 5){ 32.294}}
\put(745,270){\line(-3, 5){ 32.471}}
\put(945,536){\line(-3,-5){ 32.735}}
\put(913,481){\line( 3,-5){ 32.471}}
\put(910,481){\line(-1, 0){ 56}}
\put(224,536){\line(-3,-5){ 32.735}}
\put(192,481){\line(-1, 0){ 63}}
\put(194,478){\line( 3,-5){ 28.588}}
\put(149,480){\vector( 1, 0){ 27}}
\put(201,492){\vector( 1, 2){ 12.200}}
\put(248,480){\line( 1, 0){ 64}}
\put(312,480){\line( 3, 5){ 32.294}}
\put(314,478){\line( 3,-5){ 29.029}}
\put(333,518){\vector(-1,-2){ 12}}
\put(298,480){\vector(-1, 0){ 26}}
\put(464,426){\line(-3, 5){ 32.735}}
\put(432,481){\line(-1, 0){ 63}}
\put(434,484){\line( 3, 5){ 28.588}}
\put(389,482){\vector( 1, 0){ 27}}
\put(441,470){\vector( 1,-2){ 12.200}}
\put(488,481){\line( 1, 0){ 64}}
\put(552,481){\line( 3,-5){ 32.294}}
\put(554,483){\line( 3, 5){ 29.029}}
\put(573,443){\vector(-1, 2){ 12}}
\put(538,481){\vector(-1, 0){ 26}}
\put(705,537){\line(-3,-5){ 32.735}}
\put(673,482){\line( 3,-5){ 32.471}}
\put(670,482){\line(-1, 0){ 56}}
\put(696,445){\vector(-2, 3){ 14.923}}
\put(678,495){\vector( 2, 3){ 14.923}}
\put(823,426){\line(-3, 5){ 32.735}}
\put(791,481){\line( 3, 5){ 32.471}}
\put(788,481){\line(-1, 0){ 56}}
\put(814,518){\vector(-2,-3){ 14.923}}
\put(796,468){\vector( 2,-3){ 14.923}}
\put(251,615){\line( 3, 5){ 32.735}}
\put(283,670){\line(-3, 5){ 32.471}}
\put(286,670){\line( 1, 0){ 56}}
\put(260,707){\vector( 2,-3){ 14.923}}
\put(278,657){\vector(-2,-3){ 14.923}}
\put(369,726){\line( 3,-5){ 32.735}}
\put(401,671){\line(-3,-5){ 32.471}}
\put(404,671){\line( 1, 0){ 56}}
\put(378,634){\vector( 2, 3){ 14.923}}
\put(396,684){\vector(-2, 3){ 14.923}}
\put(586,670){\line(-1, 0){ 64}}
\put(522,670){\line(-3,-5){ 32.294}}
\put(520,672){\line(-3, 5){ 29.029}}
\put(501,632){\vector( 1, 2){ 12}}
\put(536,670){\vector( 1, 0){ 26}}
\put(610,615){\line( 3, 5){ 32.735}}
\put(642,670){\line( 1, 0){ 63}}
\put(640,673){\line(-3, 5){ 28.588}}
\put(685,671){\vector(-1, 0){ 27}}
\put(633,659){\vector(-1,-2){ 12.200}}
\put(826,669){\line(-1, 0){ 64}}
\put(762,669){\line(-3, 5){ 32.294}}
\put(760,667){\line(-3,-5){ 29.029}}
\put(741,707){\vector( 1,-2){ 12}}
\put(776,669){\vector( 1, 0){ 26}}
\put(850,725){\line( 3,-5){ 32.735}}
\put(882,670){\line( 1, 0){ 63}}
\put(880,667){\line(-3,-5){ 28.588}}
\put(925,669){\vector(-1, 0){ 27}}
\put(873,681){\vector(-1, 2){ 12.200}}
\put(592,324){\line( 3,-5){ 32.735}}
\put(624,269){\line(-3,-5){ 32.471}}
\put(601,232){\vector( 2, 3){ 14.923}}
\put(619,282){\vector(-2, 3){ 14.923}}
\put(835,215){\line( 3, 5){ 32.735}}
\put(867,270){\line(-3, 5){ 32.471}}
\put(844,307){\vector( 2,-3){ 14.923}}
\put(862,257){\vector(-2,-3){ 14.923}}
\put(458,215){\line(-3, 5){ 32.735}}
\put(426,270){\line( 3, 5){ 32.471}}
\put(449,307){\vector(-2,-3){ 14.923}}
\put(431,257){\vector( 2,-3){ 14.923}}
\put(214,326){\line(-3,-5){ 32.735}}
\put(182,271){\line( 3,-5){ 32.471}}
\put(205,234){\vector(-2, 3){ 14.923}}
\put(187,284){\vector( 2, 3){ 14.923}}
\put(177,570){\makebox(0,0){\Large $t_b$}}
\put(296,570){\makebox(0,0){\Large $s^{-1}$}}
\put(413,570){\makebox(0,0){\Large $1$}}
\put(533,570){\makebox(0,0){\Large $1$}}
\put(654,570){\makebox(0,0){\Large $1$}}
\put(774,570){\makebox(0,0){\Large $1$}}
\put(894,570){\makebox(0,0){\Large $1$}}
\put(173,370){\makebox(0,0){\Large $1$}}
\put(293,370){\makebox(0,0){\Large $1$}}
\put(413,370){\makebox(0,0){\Large $1$}}
\put(533,370){\makebox(0,0){\Large $1$}}
\put(654,370){\makebox(0,0){\Large $1$}}
\put(776,370){\makebox(0,0){\Large $s$}}
\put(898,370){\makebox(0,0){\Large $t_b$}}
\put(193,150){\makebox(0,0){\Large $1$}}
\put(317,150){\makebox(0,0){\Large $t_s$}}
\put(436,150){\makebox(0,0){\Large $s$}}
\put(847,150){\makebox(0,0){\Large $s^{-1}$}}
\put(729,150){\makebox(0,0){\Large $t_s$}}
\put(603,150){\makebox(0,0){\Large $1$}}
\put( -23,510){\makebox(0,0)[lb]{\smash{\LARGE (a)}}}
\put( -23,263){\makebox(0,0)[lb]{\smash{\LARGE (b)}}}
\end{picture}
\end{center}
\vskip 2cm

\caption{The allowed arrow configurations and corresponding
Boltzmann weights for (a) bulk and (b) surface vertices.}
\end{figure}
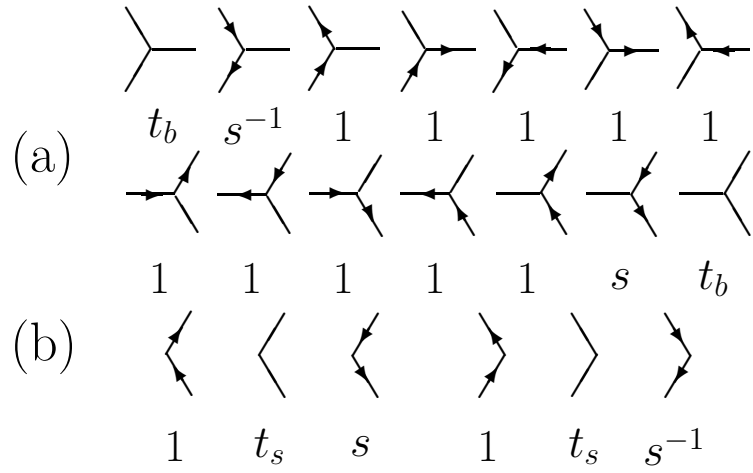


\begin{references}
\bibitem{foot1} We note that another orientation of the honeycomb
strip geometry can also be treated along the present lines.
\bibitem{foot2} These correspond to reflection matrices for which the
corresponding spin chain is not quantum group invariant and thus were
seemingly of little interest.
\end{references}
\end{document}